\documentclass{webofc}
\usepackage[varg]{txfonts}   
\usepackage{amssymb}
\usepackage{amsmath}
\usepackage{amscd}
\usepackage{latexsym}

\usepackage{graphicx}
\usepackage{array}

\newcommand{\be}{\begin{equation}}
\newcommand{\ee}{\end{equation}}
\newcommand{\Dlt}{\Delta}
\newcommand{\dlt}{\delta}
\newcommand{\prt}{\partial}

\newcommand{\bt}{\beta}
\newcommand{\vp}{\varphi}
\newcommand{\ep}{\varepsilon}
\newcommand{\al}{\alpha}
\newcommand{\ra}{\rightarrow}

\newcommand{\gm}{\gamma}
\newcommand{\om}{\omega}

\newcommand{\lbd}{\lambda}

\newcommand{\rgl}{\rangle}
\newcommand{\lgl}{\langle}

\begin{document}

\title{Describing phase transitions in field theory by self-similar approximants}

\author{\firstname{V.I.} 
\lastname{Yukalov}\inst{1,2}\fnsep\thanks{\email{yukalov@theor.jinr.ru}} 
\and
\firstname{E.P.} \lastname{Yukalova}\inst{3}
}

\institute{Bogolubov Laboratory of Theoretical Physics, 
Joint Institute for Nuclear Research, Dubna 141980, Russia
\and
     Instituto de Fisica de S\~ao Carlos, Universidade de S\~ao Paulo, \\
CP 369,  S\~ao Carlos 13560-970, S\~ao Paulo, Brazil       
\and
     Laboratory of Information Technologies, 
Joint Institute for Nuclear Research, Dubna 141980, Russia      
          }

\abstract{
 Self-similar approximation theory is shown to be a powerful tool for describing 
phase transitions in quantum field theory. Self-similar approximants present the 
extrapolation of asymptotic series in powers of small variables to the arbitrary 
values of the latter, including the variables tending to infinity. The approach 
is illustrated by considering three problems: (i) The influence of the coupling 
parameter strength on the critical temperature of the $O(N)$-symmetric 
multicomponent field theory. (ii) The calculation of critical exponents for 
the phase transition in the $O(N)$-symmetric field theory. (iii) The evaluation 
of deconfinement temperature in quantum chromodynamics. The results are in good 
agreement with the available numerical calculations, such as Monte Carlo 
simulations, Pad\'e-Borel summation, and lattice data. 
}

\maketitle

\section{Introduction}

Phase transitions in field theory are known to be a very interesting physical 
problem, whose description usually confronts complicated calculational challenges 
(see, e.g., \cite{Satz_1,Hagedorn_2,Cleymans_3,Reeves_4,Kleinert_5}). Our aim in 
this report is to show that the description of phase transitions can be efficiently 
done by an original technique called {\it self-similar approximation theory}. This 
theory allows us to find analytical expressions for the sought solutions, it is 
rather simple, involving only low-cost calculations, and at the same time, it is 
very accurate, being comparable in accuracy with heavy numerical calculations.      

The typical problem is that, as a rule, in the vicinity of a phase transition there 
are no small parameters, while calculations could be accomplished by means of 
perturbation theory in powers of such parameters, which leads to divergent series. 
There exist techniques allowing for effective summation of such series, for instance, 
Pad\'{e} approximation \cite{Baker_6} or Borel summation \cite{Kleinert_5}. However, 
these techniques not always are applicable, as is discussed in Refs. 
\cite{Baker_7,Yukalov_8,Gluzman_9,Yukalov_10}. 

Below we show that self-similar approximation theory overcomes the problems of 
asymptotic perturbation theory, making it possible to find analytic approximate 
solutions that are valid for the whole range of variables between zero to infinity. 
This theory combines simplicity with good accuracy. 

The layout of the paper is as follows: In Sec. 2, we formulate the mathematical 
problem, giving the sketch of the main ideas the self-similar approximation theory 
is based on, and present two types of resulting approximants. In Sec. 3, we 
illustrate the use of the approach for studying the influence of the coupling 
parameter strength on the critical temperature of the $O(N)$-symmetric multicomponent 
field theory. Section 4 shows how the critical exponents for the phase transition in 
the $O(N)$-symmetric field theory can be calculated. And in Sec. 5, we demonstrate 
that, being based on weak-coupling high-temperature expansions, with employing 
self-similar approximation theory, it is possible to estimate the temperature of 
deconfinement phase transition in quantum chromodynamics. Section 6 summarizes the 
results.

\section{Main ideas of self-similar approximation theory}

Suppose we are looking for a physical quantity corresponding to a real function 
$f(x)$ satisfying very complicated equations that can be solved only by means of 
perturbation theory in powers of the asymptotically small real variable $x$, 
yielding
\be
\label{1}
 f(x) \simeq f_k(x) \qquad ( x\ra 0 ) \;  ,
\ee
with the $k$-th order series
\be
\label{2}
 f_k(x) = f_0(x) \left( 1 + \sum_{n=1}^k a_n x^n \right) \;  ,
\ee
where $f_0(x)$ is a given function. At the same time, our aim is to find the values 
of the solution $f(x)$ at finite $x$ or in some cases even at $x\ra\infty$. This 
implies that we need to extrapolate the asymptotic series (\ref{2}), derived for 
small $x \ra 0$, to the whole domain of the variable $x \in [0, \infty)$. A general 
method of realizing such an extrapolation is provided by self-similar approximation 
theory \cite{Yukalov_11,Yukalov_12,Yukalov_13,Yukalov_14,Yukalov_15}, whose main 
ideas are as follows.     

\begin{enumerate}
\item
The sequence $\{f_k(x)\}$, with $k = 1,2,\ldots$ has to be reorganized 
into another sequence $\{F_k(x,u_k)\}$ by introducing parameters $u_k$ that are 
called {\it control parameters} because of their role in regulating the properties 
of the reorganized sequence $\{F_k(x,u_k)\}$. 

\item
The control parameters $u_k$ are to be converted into control functions 
$u_k(x)$ such that to force the sequence $\{F_k(x,u_k(x))\}$ becoming convergent. 

\item
It is luring to discover a law, according to which a term $F_k$ transforms into 
$F_{k+1}$. If such a transformation law is known, then it would be straightforward 
to follow the chain of transformations $F_k\ra F_{k+1}\ra\ldots\ra F^*$ leading to 
the sequence limit $F^*$ representing the sought function $f(x)$. 
\end{enumerate}

The introduction of control parameters can be done in several ways
\cite{Yukalov_11,Yukalov_12,Yukalov_13,Yukalov_14,Yukalov_15,Yukalov_16,Yukalov_17}. 
For example, they can be introduced through initial conditions. Say, we are 
considering a system characterized by a Hamiltonian (or Lagrangian) $H$. We can 
take a simpler Hamiltonian $H_0(u)$ modeling the considered system and containing 
control parameters $u$. Introducing the Hamiltonian
\be
\label{3}
 H_\ep = H_0(u) + \ep [ H - H_0(u) ] \;  ,
\ee
one can resort to perturbation theory with respect to $\varepsilon$, calculating 
wave functions, or Green functions, or physical quantities
\be
\label{4}
 F_k(x,u_k) = \lgl \hat A(x) \rgl_k 
\ee
related to some operators of observables, setting at the end $\ep\ra 1$.     
 
The other way of introducing control parameters is by reexpansion tricks. Thus, 
when $f_k(x,m)$ in (\ref{2}) depends on some parameter $m$, it is possible to make 
a substitution in (\ref{2}), such as
$$
 m \longrightarrow u + \ep ( m - u)\; , \qquad x \longrightarrow \ep x \;  ,
$$
containing a control parameter $u$, and to reexpand the resulting 
$f_k(\ep x,u+\ep(m-u))$ in powers of $\ep$, setting at the end  $\ep\ra 1$.  

Or one can incorporate a control parameter $u$ through a change of the variable 
$x=x(z,u)$ then reexpanding $f_k(x(z,u))$ in powers of $z$. 

As an example of a change of the variable, it is possible to mention the conformal 
mapping
$$
 x = \frac{4u^2z}{(1-z)^2} \; , \qquad 
z = \frac{\sqrt{x+u^2}-u}{\sqrt{x+u^2}+u} \;  .
$$

One more method of introducing control parameters is by a functional transformation
\be
\label{5}
\hat T[u_k]f_k(x) = F_k(x,u_k) \; , \qquad f_k(x) = \hat T^{-1}[u_k] F_k(x,u_k) \; ,
\ee
containing these parameters. 

One of the simplest transformations is given by the fractal transform
\be
\label{6}
 \hat T[u]f_k(x) = x^u f_k(x)   
\ee
that is used for deriving self-similar approximants. 

After control parameters are incorporated into the sequence, it is necessary to 
convert them into control functions governing the sequence convergence 
\cite{Yukalov_16,Yukalov_17}. The basic underlying idea is to connect the choice 
of control functions with the property of convergence that is expressed in the 
Cauchy criterion of convergence. The Cauchy criterion tells us that the sequence 
$\{F_k(x,u_k)\}$ converges if and only if for each $\ep > 0$ there exists such 
$n_\ep$ that
\be
\label{7}
| \; F_{k+p}(x,u_{k+p}) -F_k(x,u_k) \; | < \ep
\ee
for all $k >n_\ep$ and $p \geq 0$. 

The Cauchy criterion and the methods of optimal control theory suggest that, in 
order to induce fastest convergence, control functions have to minimize the 
fastest-convergence cost functional
\be 
\label{8}
 {\cal C}[u] = \sum_k  | \; F_{k+p}(x,u_{k+p}) - F_k(x,u_k) \; | .
\ee
Thus control functions are defined as the minimizers of this cost functional,
\be
\label{9}
 \min_u {\cal C}[u] \rightarrow u_k(x) \;  .
\ee
The practical methods of the minimization can be found in Refs. 
\cite{Yukalov_11,Yukalov_12,Yukalov_13,Yukalov_14,Yukalov_15,Yukalov_16,Yukalov_17}. 

Substituting the found control functions into $F_k(x,u_k)$ yields the {\it optimized
approximants}
\be
\label{10}
 \widetilde f_k(x) = F_k(x,u_k(x) ) \; .
\ee
These approximants already can extrapolate the asymptotic series (\ref{2}) to 
finite values of the variable $x$. But we want to go a step further, trying to find 
a transformation law between the terms $\widetilde{f}_k(x)$ and $\widetilde{f}_{k+1}(x)$, 
or more generally, a law transforming $\widetilde{f}_k(x)$ into $\widetilde{f}_{k+p}(x)$.   

To derive a transformation law between $\widetilde{f}_k(x)$ and $\widetilde{f}_{k+p}(x)$, 
we can treat this transformation as a motion in the space of approximants, with the 
approximation order playing the role of discrete time. In other words, we need to 
construct a dynamical system describing this motion. The correct mathematical description 
of a dynamical system requires to formulate the motion in terms of endomorphisms 
\cite{Hirsch_18,Crutchfield_19,Cook_20}. For this purpose, we define the expansion 
function $x = x_k(f)$ given by the reonomic constraint
\be
\label{11}
 F_0(x,u_k(x) ) = f \; , \qquad x = x_k(f) \; .
\ee
And the required endomorphism takes the form
\be
\label{12}
  y_k(f) \equiv \widetilde f_k(x_k(f) ) \; .
\ee
 
By this construction, if the sequence of the approximants $\{\widetilde{f}_k(x)\}$, 
with increasing $k$, tends to a limit $\widetilde{f}(x)$, then the sequence of the 
endomorphisms $\{y_k(f)\}$ tends to a fixed point $y^*(f)$. In the vicinity of a 
fixed point, the endomorphism satisfies the self-similar relation 
\be
\label{13}
  y_{k+p}(f) = y_k(y_p(f) ) \; .
\ee

Equation (\ref{13}) describes the motion in the space of approximants with respect 
to the discrete time $k$ being the approximation order. A dynamical system in discrete 
time is called cascade. Since it describes the motion in the space of approximants, 
it is named {\it approximation cascade}. By construction, the trajectory of the 
approximation cascade $\{y_k(f)\}$ is bijective to the sequence of approximants 
$\{\widetilde{f}_k(x)\}$, so that the fixed point $y^*(f)$ is bijective to the sequence 
limit $\tilde{f}(x)$. In that way, to get the effective limit of the sought function, 
we need to find the fixed point of the approximation cascade.

Usually, it is more convenient to deal with the equations of motion in continuous time. 
To this end, it is possible to embed the approximation cascade into an approximation 
flow that is a dynamical system in continuous time. The embedding of the cascade into 
a flow,
\be
\label{14}
 \{ y_k(f) : \; k = 0,1,2,\ldots \} \subset \{ y(t,f): t \geq 0 \} \;  ,
\ee
implies that the flow satisfies the same equation of motion
\be
\label{15}
 y(t+t',f) = y(t,y(t',f) ) \;  ,
\ee    
and the flow trajectory passes through all the points of the cascade trajectory,
\be
\label{16}
 y(k,f) = y_k(f) \;  .
\ee
      
Relation (\ref{15}) can be rewritten as the differential Lie equation
\be
\label{17}
 \frac{\prt}{\prt t}\; y(t,f) = v( y(t,f) ) \;  ,
\ee
in which the velocity $v(y)$ is analogous to the Gell-Mann-Low function in 
renormalization group approach. Integrating the above equation gives the integral 
form
\be
\label{18}
 \int_{y_k}^{y_k^*} \frac{dy}{v(y)} = t_k \;  ,
\ee
which allows us, starting from a point $y_k(f)$, and moving during time $t_k$ 
with the velocity $v_k(y)$, to find the approximate fixed point $y_k^*(f)$. Hence, 
we can find the {\it self-similar approximant} $f_k^*(x)$ for the sought function.   

Applying the procedure, described above, to the asymptotic expansion (\ref{2}) it 
is possible to derive several forms of self-similar approximants. One is the {\it 
self-similar factor approximant} \cite{Yukalov_21,Gluzman_22,Yukalov_23}
\be
\label{19}
 f_k^*(x) = f_0(x) \prod_{i=1}^{N_k} ( 1 + A_i x)^{n_i} \;  ,
\ee
where
\begin{eqnarray}
\nonumber
N_k = \left\{ \begin{array}{ll}
k/2 \; , ~ & ~ k = 2,4,\ldots \\
(k+1)/2 \; , ~ & ~ k = 3,5,\ldots 
\end{array} \right. \; .
\end{eqnarray}
This approximant provides the extrapolation of the asymptotic series (\ref{2}), 
valid only for asymptotically small $x \ra 0$, to the arbitrary values of the 
variable $x \in [0, \infty)$. All parameters $A_i$ and $n_i$, playing the role of 
control parameters, can be defined by the accuracy-through-order procedure comparing 
the asymptotic forms
\be
\label{20}
  f_k^*(x) \simeq f_k(x) \qquad ( x \ra 0 ) \; .
\ee
  
The other type of the approximants is the {\it self-similar exponential approximant}
\cite{Yukalov_24}
\be 
\label{21}
 f_k^*(x) = f_0(x) \exp( C_1 x \exp( C_2 x \ldots \exp(C_k t_k x))) \;  ,
\ee
in which 
$$
C_j = \frac{a_j}{a_{j-1}} \qquad ( j = 1,2,\ldots, k )
$$
and the control function $t_k = t_k(x)$ is defined by the equation
\be
\label{22}
 t_k = \exp(C_k x t_k ) \;  .
\ee

The choice of the type of an approximant is dictated by the physics of the treated 
problem. When the considered physical quantities are expected to vary sufficiently 
slowly with the variation of $x$, one should use the factor approximants. And when 
one expects a fast variation of the physical quantities, exponential approximants are 
more appropriate.

\section{Influence of coupling-parameter strength on critical temperature}

As an application of the self-similar approximation theory, let us consider the 
second-order phase transition in the $O(N)$-symmetric field theory in three dimensions. 
The critical temperature depends on the coupling-parameter strength 
\be
\label{23}
  \gm \equiv \rho^{1/3} a_s \; ,
\ee
where $\rho$ is average density and $a_s$, scattering length. For the free field 
theory, where $\gamma \ra 0$, the critical temperature is
\be
\label{24}
 T_0 = \frac{2\pi}{m} \left[ \frac{\rho}{\zeta(3/2)} \right]^{2/3} \;  .
\ee
The question is: how the critical temperature $T_c(\gamma)$ changes when the 
interaction is switched on?  

One considers the relative critical temperature shift
\be
\label{25}
 \frac{\Dlt T_c}{T_0} \equiv \frac{T_c(\gm) - T_0}{T_0} \; .
\ee
At small $\gamma$, the shift behaves \cite{Baym_25,Baym_26} as 
\be
\label{26}
 \frac{\Dlt T_c}{T_0} \simeq c_1 \gm \qquad ( \gm \ra 0 ) \;  .
\ee
The coefficient $c_1$ can be calculated by using the loop expansion 
\cite{Kastening_27,Kastening_28,Kastening_29} resulting in the asymptotic series 
\be
\label{27}
c_1(x) \simeq \sum_{n=1}^5 a_n x^n \qquad ( x \ra 0 )
\ee
in powers of the variable
$$
 x = (N+2)\; \frac{\lbd}{\sqrt{\mu} } \;  ,
$$
where $\lambda$ is a renormalized coupling and $\mu$, effective chemical potential. 
But the problem is that at the critical temperature, the chemical potential tends to 
zero, $\mu \ra 0$.
 
Therefore the expansion variable tends to infinity, $x \ra \infty$. So that to get 
$c_1$, we need to define $c_1(\infty)$, when expansion (\ref{27}) becomes senseless. 
However in our approach, it is possible to define the effective limit $c_1(\infty)$ 
of expression (\ref{27}) under $x \ra \infty$.    

Applying to the asymptotic series (\ref{27}) our approach, we use self-similar factor 
approximants \cite{Yukalov_30}, as is explained in the previous section, getting for 
$c_1(x)$ the approximants
\be
\label{28}  
 c_1(x) \ra f_k^*(x) = a_1 x \prod_{i=1}^{N_k} ( 1 + A_i x)^{n_i} \;  .
\ee
At large $x$, we have
\be
\label{29}
 f_k^*(x) \simeq B_k x^{\bt_k} \qquad ( x \ra \infty ) \;  ,
\ee
where
$$
B_k = a_1 \prod_{i=1}^{N_k} A_i^{n_i} \; , \qquad 
\bt_k = 1 + \sum_{i=1}^{N_k} n_i \; .
$$
The large-variable limit is finite, provided that $\beta_k = 0$, which yields the 
value
\be
\label{30}
  f_k^*(\infty) = B_k \ra c_1(\infty) 
\ee
giving the sought coefficient $c_1$. The found coefficients $c_1$ for a different 
number of components $N$ in the $O(N)$-symmetric $\varphi^4$ field theory in $3d$ 
are shown in Table 1, where they are compared with available Monte Carlo simulations 
\cite{Kashurnikov_31,Arnold_32,Arnold_33,Sun_34}.  

\begin{table}
\centering
\caption{ The coefficient $c_1$  of the critical temperature shift, for a different 
number of the field components $N$, given by self-similar factor approximants, compared 
with the available Monte Carlo simulations.}
\label{tab-1}
\begin{tabular}{|c|c|c|} \hline
$N$ &     $c_1$       &   Monte Carlo \\ \hline
0   &  0.77$\pm$ 0.03 &        \\ \hline
1   &  1.06$\pm$ 0.05 &  1.09$\pm$ 0.09   \\ \hline
2   &  1.29$\pm$ 0.07 &  1.29$\pm$ 0.05    \\ 
    &                 &  1.32$\pm$ 0.02  \\ \hline
3   &  1.46$\pm$ 0.08 &                   \\ \hline
4   &  1.60$\pm$ 0.09 & 1.60$\pm$ 0.10  \\ \hline
\end{tabular}
\end{table}

\section{Calculation of critical exponents in $\varphi^4$ field theory}    

It is also possible to calculate critical exponents for the $O(N)$-symmetric 
$\vp^4$ theory in $3d$ by applying self-similar approximation theory to the 
Wilson $\varepsilon$ expansions
\be
\label{31}
 f_k(\ep) = \sum_{n=0}^k c_n \ep^n \;  ,
\ee
where $\varepsilon = 4 - d$. Such expansions are derived for $\ep\ra 0$, while 
in reality $\ep=1$. 

The extrapolation of these asymptotic expansions can again be done by means of 
self-similar factor approximants
\be
\label{32}
  f_k^*(\ep) = f_0(\ep)\prod_{i=1}^{N_k} (1 +  A_i\ep )^{n_i} \; .
\ee
Setting here $\varepsilon = 1$, we define the answer as
\be
\label{33}
  f_k^* = \frac{1}{2} \; \left[ f_k^*(1) + f_{k-1}^*(1) \right] \; ,
\ee
with the error bar
$$
 \pm \; \frac{1}{2} \; \left[ f_k^*(1) - f_{k-1}^*(1) \right] \;  .
$$

As an example, let us consider the $O(1)$ universality class for $3d$, which 
includes such a well known case as the three-dimensional Ising model. The 
$\ep$-expansions for the critical exponents $\eta$, $\nu$, and $\omega$ have 
the form \cite{Kleinert_35,Kleinert_36}
$$
\eta \simeq 0.0185185 \ep^2 + 0.01869\ep^3 - 0.00832877\ep^4 + 
0.0256565 \ep^5 \; ,
$$
$$
\nu^{-1} \simeq 2 - 0.333333 \ep - 0.117284 \ep^2 +  0.124527 \ep^3  - 
 0.30685\ep^4 - 0.95124\ep^5 \; ,
$$
\be
\label{34}
\omega \simeq  \ep - 0.62963\ep^2 + 1.61822 \ep^3 - 5.23514 \ep^4 + 
20.7498 \ep^5 \; .
\ee
Other exponents can be found from the scaling relations
\be
\label{35}
\al = 2 - 3\nu \; , \qquad  \bt = \frac{\nu}{2}\; ( 1 + \eta ) \; , 
\qquad
\gm = \nu ( 2 -\eta) \; , \qquad \dlt = \frac{5-\eta}{1+\eta} \;   .
\ee
As is easy to check, the direct substitution of $\varepsilon = 1$ in these 
expressions leads to bad results having little to do with the quantities 
that can be obtained in experiments or in numerical calculations. While 
extrapolating these expansions by means of self-similar factor approximants 
(\ref{32}) and setting $\varepsilon = 1$, we come to the values (\ref{33}) 
that are close to those found in numerical calculations, such as the conformal 
bootstrap conjecture\cite{El_37,El_38,Gliozzi_39,Komargodski_40,Kos_41} and 
Monte Carlo simulations 
\cite{Blote_42,Janke_43,Ferrenberg_44,Baillie_45,Holm_46,Holm_47,Chen_48,
Holm_49,Li_50,Kanaya_51,Ballesteros_52,Caracciolo_53,Landau_54,Hasenbusch_55,
Campostrini_56,Pelissetto_57,Deng_58,Hasenbusch_59,Campostrini_60,Hasenbusch_61,
Ferrenberg_62}. 
Table 2 demonstrates the results obtained using the self-similar factor 
approximants, as compared with the numerical data of the conformal bootstrap 
conjecture, and Monte Carlo simulations.

\begin{table}
\centering
\caption{Critical exponents for $O(1)$-symmetric $\vp^4$ field theory in $3d$, 
calculated using self-similar factor approximants (FA), conformal bootstrap 
conjecture (CB), and Monte Carlo simulations (MC).}
\label{tab-2}
\begin{tabular}{|c|c|c|c|} \hline
        &   $FA$   &   $CB$   &   $MC$    \\ \hline
$\al$   &  0.10645 &  0.11008 &  0.11026  \\ \hline
$\bt$   &  0.32619 &  0.32642 &  0.32630  \\ \hline
$\gm$   &  1.24117 &  1.23708 &  1.23708  \\ \hline
$\dlt$  &  4.80502 &  4.78984 &  4.79091  \\ \hline
$\eta$  &  0.03359 &  0.03630 &  0.03611  \\ \hline
$\nu$   &  0.63118 &  0.62997 &  0.62991  \\ \hline
$\om$   &  0.78755 &  0.82966 &  0.830         \\ \hline
\end{tabular}
\end{table}

We have also calculated the critical exponents for other $O(N)$-symmetric 
$\vp^4$ theories, varying the number of components from $N=-2$ up to $N=10000$, 
and obtaining the values close to those derived by numerical methods, when these 
are available. It is important to stress that the critical exponents for $N=-2$ 
and $N \ra \infty$ are known exactly and that in our approach we obtain the same 
exact values shown in Table 3.     

\begin{table}
\centering
\caption{Exact critical exponents for $O(N)$-symmetric $\vp^4$ field theory in  
$3d$ for $N=-2$ and $N=\infty$.}
\label{tab-3}
\begin{tabular}{|c|c|c|} \hline
        &   $N=-2$ &  $N\ra\infty$      \\ \hline
$\al$   &  0.5  &  $-$1   \\ \hline
$\bt$   &  0.25 &  0.5    \\ \hline
$\gm$   &  1    &  2   \\ \hline
$\dlt$  &  5    &  5   \\ \hline
$\eta$  &  0    &  0   \\ \hline
$\nu$   &  0.5  &  1   \\ \hline
$\om$   &  0.8  &  1    \\ \hline
\end{tabular}
\end{table}

\section{Estimation of deconfinement temperature in quantum chromodynamics}

Self-similar approximation theory makes it possible to extrapolate asymptotic 
series at small variables to the whole domain of the latter. A very important 
question is whether it is feasible to predict the existence of a phase transition 
considering expansions very far from the transition point. Suppose we study the 
region of asymptotically weak coupling in QCD corresponding to high temperature. 
Is it possible to extract from such asymptotic expansions the information on the 
existence of the confinement-deconfinement phase transition? Below we show that 
extrapolating asymptotic series by self-similar approximants allows us to predict 
the deconfinement phase transition and to correctly estimate the transition 
temperature. 

We consider the $SU(N_c)$ QCD with three colours, $N_c=3$, and with massless 
quarks of $n_f$ flavors in the fundamental representation, with zero chemical 
potential. Using dimensional regularization and the modified minimal subtraction 
scheme $\overline{MS}$ one gets \cite{Zhai_63,Braaten_64,Kraemmer_65} the weak-coupling 
expansion of pressure in powers of the quantum chromodynamic coupling $\alpha_s$ or 
in powers of the coupling parameter $g$ connected by the relation
\be
\label{36}
 \al_s = \frac{g^2}{4\pi} \;  .
\ee
This expansion reads as
\be
\label{37}
 P(g) \simeq \frac{8\pi^2}{45}\; T^4 \; \left( \sum_{n=0}^5 c_n g^n + 
c_4' g^4 \ln g \right) \;  ,
\ee
where $c_1 = 0$ and other coefficients can be found in Refs. 
\cite{Zhai_63,Braaten_64,Kraemmer_65}. In the Stefan-Boltzmann limit, one has
\be
\label{38}
 P_0 \equiv P(0) = \frac{8\pi^2}{45}\; T^4 \; \left( 1 + 
\frac{21}{32}\; n_f \right) \;  .
\ee
It is convenient to define the relative pressure
\be
\label{39}
p_k \equiv \frac{P_k}{P_0} =  p_k(g,\mu,T) 
\ee
that is a function of the coupling $g$, renormalization scale $\mu$, and temperature 
$T$, and where $P_k$ is the right-hand side of (\ref{37}). Then we get
\be
\label{40} 
 p_k = 1 +  \sum_{n=2}^5 \overline c_n g^n + \overline c_4' g^4 \ln g  \;   ,
\ee
with $\overline{c}_n \equiv c_n/c_0$. 

Defining the renormalization scale $\mu = \mu(g,T)$ from the minimal-difference 
condition \cite{Yukalov_66} 
\be
\label{41}
p_4(g,\mu,T) - p_3(g,\mu,T) = 0
\ee
reduces the relative pressure (\ref{40}) to the form
\be
\label{42}
 p_5 = 1 + \overline c_2 g^2 + \overline c_3 g^3 + \overline c_5 g^5 \;   .
\ee
 
We extrapolate this expansion with the use of self-similar exponential approximants, 
as described in Sec. 2. Then we have the approximants to second order
$$
 p_2^* =\exp(C_2 g^2 ) \; , \qquad  C_2 = \frac{c_2}{c_0}  \;   ,
$$
to third order
$$
  p_3^* =\exp(C_2 g^2 t_3 )  \;   ,
$$
with
$$
 t_3 =\exp(C_3 g t_3 ) \; , \qquad   C_3 = \frac{c_3}{c_2}  \; ,
$$
and to fifth order
\be
\label{43}
 p_5^* =\exp(C_2 g^2 \exp ( C_3 g t_5 ) )  \;  ,
\ee   
where
$$
  t_5 =\exp(C_5 g^2 t_5 ) \; , \qquad  C_5 = \frac{c_5}{c_3} \;   .
$$

The running coupling satisfies the renormalization group equation 
\be
\label{44}
 \mu \; \frac{\prt g}{\prt \mu} = \bt(g) \;  .
\ee
The Gell-Mann-Low function can be found \cite{Luthe_67,Baikov_68} in the 
weak-coupling limit
$$
\bt(g) \simeq \bt_k(g) \qquad ( g \ra 0 ) 
$$
as a five-order expansion
\be
\label{45}
 \bt_k(g) = - \sum_{n=0}^k b_n g^{2n+3} = - b_0 g^3 \left ( 1 +
\sum_{n=1}^k \frac{b_n}{b_0} \; g^{2n} \right) \;  ,
\ee
with $k = 5$ and the coefficients given in Ref. \cite{Luthe_67,Baikov_68}. As 
an initial condition for Eq. (\ref{44}), we can take the value of the coupling for 
the $Z^0$ boson mass:
\be
\label{46}
 g(m_Z) = 1.22285 \qquad  ( m_Z = 91187\; {\rm Mev} ) \;  .
\ee
The self-similar extrapolation of the Gell-Mann-Low function is
\be
\label{47}
 \bt_k^*(g) = - b_0 g^3 \exp\left( B_1 g^2 \exp\left( 
B_2 g^2 \ldots \exp\left( B_k t_k g^2 \right) \right) \right) \;  ,
\ee
where
$$
B_j = \frac{b_j}{b_{j-1} } \; , \qquad t_k = \exp\left( B_k g^2 t_k\right) \;   .
$$  
Solving the renormalization group equation (\ref{44}), with the Gell-Mann-Low 
function  (\ref{47}) gives $g = g(\mu)$. Taking into account that the 
minimal-difference condition (\ref{41}) defines $\mu = \mu(g,T)$, we find the 
temperature dependence for $g = g(T)$ and $\mu = \mu(T)$. Substituting this into 
pressure (\ref{43}) gives the temperature dependence $p_5 = p_5(T)$. The behavior 
of the latter for $n_f = 6$ is shown in Fig. 1. With diminishing temperature from 
high values, the pressure, first, slightly deviates from the Stefan-Boltzmann
limit and then sharply falls down to zero at $T_c \sim 150$ MeV. This temperature, 
representing the deconfinement phase transition, is in agreement with lattice data 
\cite{Borsanyi_69,Borsanyi_70}, as well as with some statistical models 
\cite{Yukalov_71,Yukalov_72}.    

\begin{figure}[ht]
\centering
\includegraphics[width=7cm]{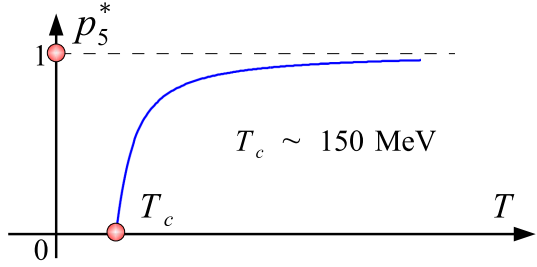}
\caption{Behavior of temperature dependence $p_5=p_5(T)$ for $n_f = 6$.}
\end{figure}

In this way, extrapolating high-temperature weak-coupling expansions by self-similar 
approximants shows the existence of deconfinement phase transition with a reasonable 
estimation of the deconfinement temperature of $T_c \sim 150$ MeV. 

\section{Conclusion}

We have shown that self-similar approximation theory is a powerful tool for 
extrapolating asymptotic series in small variables to the whole region of the latter 
from zero to infinity. The application of the approach is illustrated by studying the 
influence of the coupling parameter strength in the $O(N)$-symmetric $\vp^4$ theory 
on the critical temperature of symmetry breaking. The critical exponents for this 
phase transition are calculated. The results are in very good agreement with those 
of numerical methods, such as Monte Carlo simulations. 

Also, we show that the approach makes it possible to predict the 
confinement-deconfinement phase transition in QCD by extrapolating high-temperature 
weak-coupling expansions. The predicted deconfinement temperature is in agreement 
with lattice calculations.        
 
The important feature of the self-similar approximation theory is its simplicity, 
since it involves only low-cost calculations, as compared with numerical methods. 

Note that the use of Pad\'{e} approximants for this problem does not lead to 
reasonable results, exhibiting rather chaotic behavior with unphysical poles.


\end{document}